\documentclass[12pt]{article}
\usepackage{hyperref}
\usepackage{amsmath,amssymb}
\usepackage{graphicx}
\usepackage{color}
\definecolor{myblue}{rgb}{0.14,0.11,0.49}
\definecolor{myred}{rgb}{0.74,0.22,0.15}
\definecolor{mygreen}{rgb}{0.05,0.52,0.42}
\definecolor{myyellow}{rgb}{0.96,0.92,0.13}
\definecolor{myorange}{rgb}{1,0.61,0.36}
\definecolor{mypurple}{rgb}{0.71,0.02,1}
\definecolor{noir}{gray}{0.} 

\newcommand{\Couleur}[1]{\textcolor{noir}{#1}}

\definecolor{htc}{rgb}{1,1,1} 

\newcommand{\abs}[1]{\left\vert#1\right\vert}
\def\be{\begin{equation}}
\def\ee{\end{equation}}
\def\bea{\begin{eqnarray}}
\def\eea{\end{eqnarray}}
\def\bc{\begin{center}}
\def\ec{\end{center}}
\def\bi{\begin{itemize}}
\def\ei{\end{itemize}}
\def\bs{\begin{slide}}
\def\es{\end{slide}}
\def\dd{\mathrm{d}}
\def\iC{\mathrm{i}}
\def\noi{\noindent}
\title{An analytical model for the Maxwell radiation field in an axially symmetric galaxy}

\author{
Mayeul Arminjon\\
\small\it Univ. Grenoble Alpes, CNRS, Grenoble INP
, 3SR, F-38000 Grenoble, France\\
\small\it  E-mail: Mayeul.Arminjon@3sr-grenoble.fr
} 
\date{}
\begin{document}

\maketitle

Short title: {\it Maxwell radiation field in an axisymmetric galaxy}

\begin{abstract}

\noi The Maxwell radiation field is an essential physical characteristic of a galaxy. Here, an analytical model is built to simulate that field in an axisymmetric galaxy. This analytical model is based on an explicit representation for axisymmetric source-free Maxwell fields. In a previous work, the general applicability of this representation has been proved. The model is adjusted by fitting to it the sum of spherical radiations emitted by the composing ``stars". The huge ratio distance/wavelength needs to implement a numerical precision better than the quadruple precision. The model passes a validation test based on a spherically symmetric solution. The results for a set of ``stars'' representative of a disk galaxy indicate that the field is highest near to the disk axis, and there the axial component of ${\bf E}$ dominates over the radial one. This work will allow us in the future to check if the interaction energy predicted by an alternative theory of gravitation might be a component of dark matter. \\

\noi {\bf Keywords:} Disk galaxy; Maxwell equations; axial symmetry; exact solutions; numerical model; dark matter.\\


\end{abstract}


\section{Introduction}

Apart from pure magnetic fields, which are thought to be produced by a galactic dynamo action \cite{BeckWielebinski2013, Chamandy-et-al2013}, the electromagnetic (EM) field in a galaxy is in the form of EM radiation, i.e., of propagating EM waves covering the whole spectrum, from radio to gamma. Of course, a lot of information can be found in the literature regarding the production of the EM radiation field by stars and by other astrophysical objects; about its interaction with dust, as well as other processes of radiative transfer; or about the EM wave spectrum and its dependence on the position in the galaxy, etc.  \cite{PorterStrong2005, Maciel2013}. However, it seems that little or nothing can be found about the description of the EM radiation field in a galaxy as an exact solution of the Maxwell equations. The aim of the present work is to propose and first check a method to obtain such relevant solutions. \\

Our main motivation for that work is to make a step towards testing the following prediction \cite{A57} of an alternative theory of gravity, which is a scalar theory with a preferred reference frame: In the presence of both a variable gravitational field and an EM field, the total energy(-momentum-stress) tensor is not the sum of the energy tensors of matter and the EM field --- there must appear a specific interaction energy tensor, which should be distributed in space, and be gravitationally active. (Ref. \cite{B41} summarizes Ref. \cite{A57} while also giving the main motivations for that alternative theory of gravitation.) Essentially: if instead one assumes the additivity of the energy tensors of matter and the EM field, then charge is not conserved in a variable gravitational field according to that theory. (See Ref. \cite{A54}, which contains also a summary of that theory.) Thus one needs to introduce that interaction energy tensor, the form of which is determined by asking that it should be Lorentz-invariant in the absence of a gravitational field \cite{A57,B41}. That interaction energy could possibly contribute to the dark matter, because in addition to being gravitationally active and distributed in space, it has an ``exotic'' character, being different from both the energy of matter and that of the EM field. (It has a classical nature, however, as has the theory \cite{A57,A54}. I.e., the theory does not say anything about the possibility that the interaction energy might result from underlying quantum particles.) The interaction energy tensor is characterized by a scalar (field), which is determined by the gravitational and EM fields in the preferred frame of the theory; hence it depends  also on the velocity of the reference frame used, with respect to that preferred frame. More precisely, in order to calculate that scalar field in a weak and slowly varying gravitational field, one has to know the  EM field and its first-order derivatives, as well as the time derivative $\partial_T U$ and the spatial gradient of the latter \cite{A57}. (Here $U$ is the Newtonian potential.)\\

The precise goal of the present work, therefore, is to build a representative analytical model for the EM radiation field in a galaxy --- especially for the spatio-temporal variation of that EM field. This is in order to be able later to calculate the interaction energy predicted by the scalar theory \cite{A57}, and to check if its distribution resembles a ``dark halo''. However, the present goal: to describe the EM radiation field in a galaxy as an exact solution of the Maxwell equations, is important more generally, as that field is an essential physical characteristic of a galaxy. The work done in this paper is independent of any theory of gravity. Section \ref{ModelGeneral} describes the model that has been built in this work. Section \ref{Implementation} discusses its numerical implementation. The numerical results obtained so far are discussed in Sect. \ref{Results}. Finally, we present our \hyperref[Conclusion]{conclusions}.

\section{Description of the model}\label{ModelGeneral}

  \subsection{Main assumptions}
Let us state and comment the two essential assumptions of the model, i.e., the ones which it would be difficult to change without going to a different model:\\

\hypertarget{Ass1}{(i) {\it The structure of the interstellar radiation field is determined by the sum of spherical scalar potentials emitted by the stars.}} The relevant potentials are defined below, Eq. (\ref{psi_spher'}), and the corresponding sum is given by Eq. (\ref{sum_psi_spher_spectre}). That sum defines the source input that we use to determine the parameters of a general solution of the Maxwell equations, see the end of this paragraph. Note that this assumption is different from assuming that the total interstellar radiation field is the sum of the EM fields emitted by the stars. The latter assumption would be inappropriate, because it would mean neglecting the effect of important physical processes like dust extinction (by absorption and scattering), and, more generally, radiative transfer. Nevertheless, according to Maciel \cite{Maciel2013}, ``The interstellar radiation field in the optical and ultraviolet comes essentially from integrated stellar radiation". Note, moreover, that in directions which are roughly orthogonal to a galactic disk or at least are strongly inclined with respect to it, the light emitted by the stars of that galaxy will suffer much less alteration as compared with what happens close to the galactic plane. This is important in connection with the aim of checking if the ``interaction energy" alluded to in the Introduction might significantly contribute to the dark matter halos, since, precisely, the most part of such a halo is outside the galactic disk. Even more important in this connection is an essential feature of our model: that it provides an EM field which is an exact solution of the Maxwell equations. In contrast, if one would try to account directly for absorption, emission, and scattering processes, this feature would be quite difficult to maintain. But on the other hand, in our model, the sum of the spherical potentials generated by the point sources (by the ``stars") is fitted by a general analytical solution, see the \hyperref[Theorem]{Theorem} and see the \hyperref[Model]{statement of the model} below. The sum just mentioned thus merely serves to determine the ``shape" (the parameters) of that general solution. {\it That analytical solution, in itself, is valid independently of whether the corresponding EM field has been generated directly or has undergone various radiative transfers such as absorption, scattering, etc.}\\

\hypertarget{Ass2}{(ii) {\it The distribution of the stars and (hence) the interstellar radiation field are axially symmetric.}} This is of course not exactly true (cf. e.g. the arms of a spiral galaxy), but it seems to be a reasonable simplification which should provide a correct first approximation. Except for peculiar cases, e.g. if there were a correlation between the intensity of the EM field emitted by a star and its angular position in the galaxy, and except for the vicinity of a star, the axisymmetry of the stars' distribution should indeed imply that of the interstellar radiation field. In any case, we shall assume that both the stars' distribution and the interstellar radiation field are axially symmetric.\\



  \subsection{Distribution of the stars}\label{StarDistribution}
\bi

\item Each star (or bright object) is schematized as a point ${\bf x}$ determined by its cylindrical coordinates \Couleur{$(\rho, \phi, z)$}: distance to the symmetry axis, azimuth, altitude, thus ${\bf x} = {\bf x}(\rho, \phi, z)$.

\item A discrete set of such points, \Couleur{$\mathrm{S} = \{{\bf x}_i\} = \{{\bf x}(\rho _m, \phi _{m p q}, z_{m p}) \}$} (that is, $i = i(m,p,q)$), is got by random generation, as follows:

\bi

\item An exponential distribution is assumed for \Couleur{$\rho >0$}, i.e., $n_\rho $ values $\rho_m$ ($m=1,...,n_\rho$) are got by quasi-random generation with a probability: 
\be\label{P_h}
\Couleur{P(a<\rho_m <b)= \frac{1}{h}\int_{a} ^{b} e^{-\frac{\rho }{h}} \dd \rho },
\ee
where $h$ is the scale length, with $h =3 \,\mathrm{kpc}$ in the numerical computations. Such a distribution, with this value of $h$, is approximately correct for the Milky Way \cite{KentDameFazio1991, Robin-et-al-1992, Porcel-et-al-1998}.

\item Also, an exponential distribution is assumed for \Couleur{$z>0$}: for any $m = 1,..., n_\rho$, we draw $n_z/2$ values $z_{m p}$ ($n_z$ being an even integer) by quasi-random generation with a probability law independent of $m$:
\be\label{P_hz}
\Couleur{P(a<z_{m p} <b)= \frac{1}{h_z}\int_{a} ^{b} e^{-\frac{z }{h_z}} \dd z },
\ee
with $h_z =0.2 \,\mathrm{kpc}$ in the numerical computations. This also roughly corresponds to the Milky Way \cite{Schneider2006}.

\item For each value $z_{m p}>0$ thus obtained, we introduce another value $-z_{m p}$, i.e., we impose a perfect symmetry w.r.t. $z = 0$ in the distribution of $z$.

\item Finally, for any two $m = 1,..., n_\rho$ and $p = 1,..., n_z$, we draw $n_\phi $ values $\phi_{m p q}$, with a uniform distribution between $0$ and $2\pi$ (thus ensuring the axial symmetry of the distribution of the ``stars", as announced).

\ei
\ei


  \subsection{Explicit representation for axisymmetric EM radiation fields}

In Ref. \cite{GAZR2014}, two classes of time-harmonic axisymmetric solutions of the Maxwell equations were introduced. Although the aim of that paper \cite{GAZR2014} was to obtain in explicit form nonparaxial EM beams, these two classes are relevant for the present, very different work. The {\it first class of solutions} is got in the following way. One starts from a time-harmonic axisymmetric solution $\Psi (t,\rho ,z)=e^{-\iC \omega t} \hat{\Psi } (\rho ,z)$ of the scalar wave equation, and one associates with it a vector potential ${\bf A}$ by 
\be\label{A_from_Az}
\Couleur{{\bf A} := \Psi {\bf e}_z,\qquad \mathrm{or}\quad A_z :=\Psi , \ A_\rho  = A_\phi  =0}. 
\ee
(We shall denote by $\,({\bf e}_\rho , {\bf e}_\phi , {\bf e}_z)\,$ the standard, point-dependent, direct orthonormal basis associated with the cylindrical coordinates $\,\rho, \phi ,z\,$.) In the time-harmonic case considered for the moment, such a vector potential defines uniquely --- as it was shown in Ref \cite{GAZR2014} and, in more detail, in Ref. \cite{A60} --- the following {\it exact solution of the source-free Maxwell equations} (i.e., the Maxwell equations in a domain where the source (the 4-current) is zero 
\footnote{\
We are indeed willing to describe the EM radiation field on a galactic spatial scale, thus excluding from consideration the (much smaller) stars of the galaxy, which are the primary source of that radiation field. Given any axisymmetric solution $A_z$ of the standard wave equation in some open domain for the spacetime variables $t,\rho,z$, Eqs. (\ref{Bfi})--(\ref{Ez}) define a solution of the source-free Maxwell equations in that same subdomain \cite{GAZR2014,A60}. 
}
 ) :
\bea\label{Bfi}
\Couleur{B_\phi} & = & \Couleur{-\frac{\partial A_z}{\partial \rho }},\qquad \Couleur{E_\phi =0},\\
\nonumber\\
\label{Erho}
\Couleur{E_\rho} & = & \Couleur{\iC \frac{c^2}{\omega } \frac{\partial^2 A_z}{\partial \rho \,\partial z}},\qquad \Couleur{B_\rho =0},\\
\nonumber\\
\label{Ez}
\Couleur{E_z} & = & \Couleur{\iC \frac{c^2}{\omega } \frac{\partial^2 A_z}{\partial z^2 } + \iC \omega A_z}, \qquad \Couleur{B_z=0}.
\eea
In Ref. \cite{GAZR2014}, this class was defined only when the scalar wave $\Psi $ has the following form:
\be\label{psi_monochrom}
\Couleur{\psi _{\omega\ S} \,(t,\rho,z) = e^{-\iC \omega t} \int _{-K} ^{+K}\ J_0\left(\rho \sqrt{K^2-k^2}\right )\ e^{\iC k \, z} \,S(k)\, \dd k},
\ee
with $\omega $ the angular frequency, $K:=\omega /c$,
\footnote{\ 
In Ref. \cite{GAZR2014}, $K$ was defined as $K:=2\omega /c$ instead. 
} 
and $J_0$ the first-kind Bessel function of order \Couleur{$0$}. ($c$ is the velocity of light.) Any {\it totally propagating,} 
\footnote{\ 
This is the terminology used in Ref. \cite{GAZR2014}. We can give the following justification. Eq. (\ref{psi_monochrom}) is an integral over $k $ of ``Bessel beams", each having the form \cite{Durnin1987}
\bea\label{average_planewave}
D _{\omega \, k  \, k' } \,(t,x,y,z) & := & e^{\iC(k z - \omega t)} \int_0 ^{2 \pi} \exp[ \iC k' (x \,\cos \phi + y \,\sin \phi )] \frac{\dd \phi }{2 \pi}  \\ 
\label{Bessel_beam}
& \equiv  & e^{\iC(k z - \omega t)} \ J_0\left(k'  \rho \right )  \qquad (\rho :=\sqrt{x^2+y^2}) ,
\eea
with
\be\label{sum=K^2}
 k ^2 + k'^2 =K^2:=\omega ^2/c^2.
\ee
Eq. (\ref{average_planewave}) is the average over the angle $\phi $ of plane waves $W_\phi $ having $k_x= k' \cos \phi $, $k_y = k' \sin \phi $, and $k_z=k $, and all having the same angular frequency $\omega $. However, in Eq. (\ref{average_planewave}), $k $ and $k' $ may be complex and are only subjected to the constraint (\ref{sum=K^2}). For any such pair $(k ,k' )$, Eq. (\ref{average_planewave}) defines an axisymmetric solution of the scalar wave equation, {\it and} may be rewritten as (\ref{Bessel_beam}). In particular, in the case $k  $ real with $\abs{k  } > K$, we get $k'  = \pm \iC \sqrt{k^2 - K^2}$, so that then Eq. (\ref{average_planewave}) is an angular average of {\it ``evanescent" plane waves,} i.e., the amplitude of which depends exponentially on $x \,\cos \phi + y \,\sin \phi$. Thus, if in Eq. (\ref{psi_monochrom}) one replaces the integration bounds $\pm K$ by $\pm K'$, with $K'>K$, then, depending on whether $\abs{k  } \le K$ or $\abs{k  } > K$, $D _{\omega \, k  \, k' }$ will be an average of either propagating plane waves (with a real wave vector) or ``evanescent" ones. Whereas, leaving the bounds $\pm K$ means considering a {\it totally propagating solution}, i.e., as stated in Ref. \cite{GAZR2014}, one that does not contain ``evanescent" waves. 
} 
time-harmonic, axisymmetric solution of the scalar wave equation can be set in the form (\ref{psi_monochrom}) \cite{ZR_et_al2008, GAZR2014}. However, as noted in Ref. \cite{A60}, this form is not necessary at this stage and the solution (\ref{Bfi})--(\ref{Ez}) applies whether $A_z $ is totally propagating or not. On the other hand, given any integrable function $S(k)$ ($S\in \mathrm{L}^1([-K, +K])$, Eq. (\ref{psi_monochrom}) defines an axisymmetric solution of the standard wave equation for the whole range of values of the spacetime variables $t,\rho,z$, thus in the domain $\Omega $: ($t, z \in \mathbb{R}$, $\rho \in \mathbb{R}_+$) (i.e. $\rho \ge 0$)  \cite{ZR_et_al2008, A60}.  Therefore, given any frequency $\omega >0$ and any integrable function $S(k)$, Eqs. (\ref{Bfi})--(\ref{Ez}), applied with $A_z=\psi _{\omega\ S}$ given by Eq. (\ref{psi_monochrom}), define an axisymmetric solution of the source-free Maxwell equations in the whole spatiotemporal domain, $\Omega $.\\

The {\it second class of solutions} of the source-free Maxwell equations is deduced from the first one above by applying the EM duality to any solution of the first class, i.e., by setting \cite{GAZR2014}:
\be\label{dual}
{\bf E}'=c{\bf B}, \quad {\bf B}' = -{\bf E}/c.
\ee 

\vspace{4mm}
In the work \cite{A60}, we showed that, by combining these two classes, one can define a method that allows one to get actually {\it all} totally propagating, time-harmonic, axisymmetric source-free Maxwell fields --- and thus, by the appropriate summation on frequencies, all totally propagating axisymmetric source-free Maxwell fields. (The necessary restriction of the method to totally propagating fields turns out to be appropriate to describe the radiation field.) The main result that allows this is the following one:

\paragraph{Theorem \cite{A60}.}\label{Theorem} {\it Let $({\bf A}, {\bf E}, {\bf B})$ be any time-harmonic axisymmetric solution of the source-free Maxwell equations} (whether totally propagating or not). {\it There exist a unique solution $({\bf E}_1, {\bf B}_1)$ of the first class (\ref{Bfi})--(\ref{Ez}) and a unique solution $({\bf E}'_2, {\bf B}'_2)$ of the second class, both with the same frequency as has $({\bf A}, {\bf E}, {\bf B})$, and whose sum gives just that solution:}
\be
{\bf E} = {\bf E}_1 + {\bf E}'_2,\qquad {\bf B} = {\bf B}_1 + {\bf B}'_2.
\ee

\vspace{3mm}
Of course, as usual, it is implicit that, in Eqs. (\ref{Bfi})--(\ref{Ez}),  $B_\phi$, $E_\rho$ and $E_z$ are actually the {\it real parts} of the respective r.h.s. Therefore, when $A_z$ is totally propagating and hence has the form (\ref{psi_monochrom}), we obtain (using the fact that $\dd J_0 /\dd x = -J_1(x)$):
\be\label{Bphi_mono}
B_{\phi \,\omega \, S} = {\mathcal Re} \left[e^{-\iC  \omega t} \int_{-K} ^{+K} \sqrt{K^2-k^2}\, J_1\left(\rho \sqrt{K^2-k^2}\right ) \,S(k) \,e^{\iC kz} \dd k \right ],
\ee
\be\label{Erho_mono}
E_{\rho \, \omega \, S} = {\mathcal Re} \left[-\iC \frac{c^2}{\omega } e^{-\iC  \omega t} \int_{-K} ^{+K} \sqrt{K^2-k^2}\, J_1\left(\rho \sqrt{K^2-k^2}\right )\iC k \,S(k) \,e^{\iC kz} \dd k \right ],
\ee
\be\label{Ez_mono}
E_{z \, \omega \, S} = {\mathcal Re} \left[\iC e^{-\iC \omega t} \int_{-K} ^{+K} J_0\left(\rho \sqrt{K^2-k^2}\right )\,\left(\omega -\frac{c^2}{\omega }\,k^2 \right )\,S(k)\,e^{\iC kz} \dd k \right ],
\ee
where $K := \omega /c$. \\

As we already mentioned, the case of a general time dependence is deduced from the case with harmonic time dependence by considering a frequency spectrum. We shall consider a discrete (and finite) spectrum for simplicity. A totally propagating solution of the wave equation with a finite frequency spectrum $(\omega _j)\ (j=1,...,N_\omega )$ is got by summing solutions of the form (\ref{psi_monochrom}):
\be\label{psi_avec_spectre}
\Couleur{\Psi _{(\omega _j)\ (S_j)}\,(t,\rho,z) = \sum _{j=1} ^{N_\omega } \,\psi _{\omega_j\ S_j}\,(t,\rho,z)},
\ee
where each $\psi _{\omega_j\ S_j}$ is a time-harmonic solution having the form (\ref{psi_monochrom}). Note that the different frequencies do not necessarily have the same weight, since any given ``wave vector spectrum" $S_j$ can be multiplied by a factor $w_j$. In other words, the weights are contained in the functions $S_j$.


  \subsection{The case of spherical waves}

If, in the axisymmetric time-harmonic solution (\ref{psi_monochrom}), one puts 
\be\label{exact spectrum spherical}
\Couleur{S(k) \equiv  \frac{c}{2\omega } \quad (-K< k < K:=\frac{\omega}{c}\,)  }, 
\ee
then one gets a {\it spherical} time-harmonic solution of the scalar wave equation: This yields indeed \cite{GAZR2014,GradshteynRyzhik2007} 
\be\label{psi_spher}
\Couleur{\psi _{\omega\ S \equiv  \frac{c}{2\omega }} \ (t,\rho,z) = e^{-\iC \omega t} \,\mathrm{sinc}\left(K r\right)}, \qquad K:=\frac{\omega}{c},
\ee
where \Couleur{$ \mathrm{sinc}\, \theta  := \frac{\sin \theta }{\theta }$},\quad \Couleur{$\ r:=\abs{{\bf x}} = \sqrt{\rho ^2+z^2}$}. However, because $\sin Kr =(e^{\iC Kr}-e^{-\iC Kr})/(2\iC)$, this solution contains both an outgoing wave and an ingoing wave, and therefore it does not satisfy the Sommerfeld radiation condition. Up to a multiplying coefficient (amplitude), there is only one outgoing spherical solution of the scalar wave equation that is time-harmonic with a given frequency $\omega $, and of course it is well known:
\be\label{psi_spher'}
\psi' _\omega \ (t,\rho,z) = \frac{e^{\iC (Kr-\omega t)}} {Kr}, \qquad K:=\frac{\omega }{c}. 
\ee
Obviously, this is a totally propagating solution. Nevertheless, unfortunately, it does not seem be amenable to the integral form (\ref{psi_monochrom}) with an analytical function $S(k)$ --- in contrast to Eq. (\ref{psi_spher}). However, for a spherical {\it source} (as opposed to a sink), we must use Eq. (\ref{psi_spher'}).

\vspace{4mm}
If we have a set of spherical sources situated at the points \Couleur{${\bf x}_i$}, all sources having the same amplitude and the same frequency spectrum, (\ref{psi_spher'}) becomes: 
\be\label{sum_psi_spher_spectre}
\Psi _{({\bf x}_i)\ (\omega _j)\ (S'_j)}\,(t,\rho,z) = \sum _{i=1} ^{i_\mathrm{max}} \sum _{j=1} ^{N_\omega} \,S'_j\,\frac{e^{\iC (K_j\, r_i-\omega_j t)}} {K_j \ r_i},
\ee
where \Couleur{$r_i:=\abs{{\bf x}-{\bf x}_i}$, $K_j:=\omega _j/c$}, and setting again the initial phases to zero for simplicity. Of course one might also make the amplitude of the source, as well as the weights affected to the different frequencies $\omega _j$, depend on the source, i.e. on the index $i$, by giving a dependence on $i$ to the positive numbers $S'_j$, which would thus become $S'_{i j}$.


  \subsection{The model}\label{Model}
\bi 
\item \underline{Step {\bf 1}.} Determine a relevant axisymmetric solution $\Psi$ of the scalar wave equation,  by fitting to the form (\ref{psi_avec_spectre}) the sum (\ref{sum_psi_spher_spectre}) of the spherical radiations emitted by the ``stars" that make the ``galaxy". \\

\item \underline{Step {\bf 2}.} 
\bi
\item (2.1) Calculate the associated EM field of the first class, \Couleur{$(E_\rho, E_z, B_\phi )$} with $B_\rho = B_z = E_\phi = 0$, by summing its time-harmonic contributions given by Eqs. (\ref{Bfi})--(\ref{Ez}), with successively $A_z:=\psi _{\omega_j\ S_j} \ (j=1,...,N_\omega )$, where the functions $\psi _{\omega_j\ S_j} \ (j=1,...,N_\omega )$ are the result of step {\bf 1}.

\item (2.2) Similarly, calculate the associated  EM field of the second class, \Couleur{$(B'_\rho, B'_z, E'_\phi )$} with $E'_\rho = E'_z = B'_\phi=0$, by using the EM duality (\ref{dual}). 
\ei

\ei

Step {\bf 1} (especially) is delicate numerically, as we will see in the next section. Therefore, we shall focus in this paper on Steps 1 and (2.1). Thus, {\it in the sequel of this paper, we shall omit step (2.2), i.e., we shall consider only solutions of the first class.} This means that we shall obtain axisymmetric EM fields having $B_\rho = B_z = E_\phi =0$. ``Complete'' axisymmetric EM fields, obtained by summing solutions of the two classes, will of course have to be considered in the future work. At the stage of the fitting (Step 1), one may think to consider two different frequency spectra for the two classes, e.g. ``mutually interpenetrating" ones.



\section{Numerical implementation}\label{Implementation}

  \subsection{Precise object of the fitting}

The fitting of the sum (\ref{sum_psi_spher_spectre}) is done to determine the ``wave vector spectra" \Couleur{$S_j$} in Eq. (\ref{psi_avec_spectre}): 
\be\label{psi_avec_spectre_2}
\Couleur{\Psi _{(\omega _j)\ (S_j)}\,(t,\rho,z) = \sum _{j} \,\psi _{\omega_j\ S_j}\,(t,\rho,z)} ,
\ee
where [Eq. (\ref{psi_monochrom}) with $\omega = \omega _j$ and $K_j =\omega_j /c$]
\be\label{psi_monochrom_j}
\Couleur{\psi _{\omega_j \ S_j} \,(t,\rho,z) = e^{-\iC \omega_j t} \int _{-K_j} ^{+K_j}\ J_0\left(\rho \sqrt{K_j^2-k^2}\right )\ e^{\iC k \, z} \,S_j(k)\, \dd k}.
\ee
Thus, we have one spectrum \Couleur{$S_j$} for each value of the index \Couleur{$j=1,...N_\omega $}, the latter specifying the frequency \Couleur{$\omega _j$}. To determine these spectra, several methods could be {\it a priori} envisaged. However, a difficulty comes from the huge ratio 
\be\label{kpc/lambda}
\Couleur{\frac{\mathrm{galactic\ distances}}{\mathrm{wavelength}} \simeq \frac{\mathrm{kpc}}{ \mu \mathrm{m}} \simeq 3 \times 10^{25}},
\ee
which is the order of magnitude of the arguments of the Bessel function $J_0$ and the complex exponential in Eq. (\ref{psi_monochrom_j}). This huge number discards several possibilities. First, it turns out to be not tractable at all here to determine each \Couleur{$S_j$} by its Fourier coefficients $C_{j n}$ --- as proposed (for a very different problem) by Garay-Avenda\~no \& Zamboni-Rached \cite{GAZR2014}. Indeed, considering (to begin with) one axisymmetric time-harmonic solution (\ref{psi_monochrom}) of the scalar wave equation, this method leads to the following expansion \{Eq. (8) in Ref. \cite{GAZR2014}\}:
\be\label{expand_psi}
\psi _{\omega\ S} \,(t,\rho,z) = 2K e^{-\iC \omega t} \sum_{n=-\infty } ^\infty C_n \,\mathrm{sinc} (h_{\omega \, n}(\rho , z)),
\ee
with \,$K=\omega /c$ \,and
\be\label{h_n}
h_{\omega \, n}(\rho , z) = \sqrt{K^2 \rho ^2 +(z K + \pi n)^2}.
\ee
Introducing the wavelength $\lambda = 2\pi c/\omega = 2 \pi /K$, we have from (\ref{h_n}):
\be\label{h_n^2}
\left ( h_{\omega \, n}(\rho , z)) \right )^2 = \pi ^2 \left [ \left (\frac{2 \rho }{\lambda } \right )^2 + \left (\frac{2 z }{\lambda } + n \right )^2 \right ].
\ee
On the r.h.s. of Eq. (\ref{h_n^2}), $\rho /\lambda $ and $z/\lambda $ have the huge magnitude (\ref{kpc/lambda}). Therefore, the functions $h_{\omega \, n}$, hence also the functions $\mathrm{sinc} (h_{\omega \, n})$ which form the basis in the expansion (\ref{expand_psi}), are practically independent of $n$ for relevant values of the spatial variables $\rho $ and $z$, unless $n$ would take similarly huge values. In that case, presumably, an integer of an akin value should give the number of the different values $n$ to be taken in order to have an accurate expansion (\ref{expand_psi}) --- which of course is not tractable. Anyway, this method has been tried in this work and has not allowed us to get an accurate fitting of the sum (\ref{sum_psi_spher_spectre}). \\

In order to determine the ``spectra" $S_j$ in Eq. (\ref{psi_avec_spectre}) by fitting the sum (\ref{sum_psi_spher_spectre}) to this equation, a second method is {\it a priori} conceivable, and has indeed been tried in this work: by inverse Fourier transform. Considering again one axisymmetric time-harmonic solution (\ref{psi_monochrom}) of the scalar wave equation, and removing its dependence in $t$, a formal inverse Fourier transform gives us
\be\label{S_by_Inverse_F}
S(k) = \frac{1}{2\pi J_0\left(\rho \sqrt{K^2-k^2}\right )} \,\int_{-\infty } ^{+\infty } \psi (\rho ,z) e^{-\iC k z} \dd z.
\ee
This should thus be independent of $\rho $ when $e^{-\iC \omega t} \psi(\rho ,z) $ is indeed an axisymmetric time-harmonic solution, with frequency $\omega $, of the wave equation. What we found numerically using the Matlab software is that, for the solution (\ref{psi_spher}) corresponding to a spherical wave, with the frequency $\omega _0$ being defined in Sect. \ref{Results} below: i) The spectrum $S$ obtained by Eq. (\ref{S_by_Inverse_F}) depends on $\rho $, which it shouldn't. ii) For a given value of $\rho $, $S(k)$ varies with $k \in ]-K, +K[$ instead of being the constant $S(k) \equiv  \frac{c}{2\omega }= \frac{1}{2K }$. iii) For values of $\rho $ in the investigated range ($10^{-3}\, \mathrm{kpc},..., 10^{2}\, \mathrm{kpc}$), $S(k)$ is much smaller than that value $\frac{1}{2K }$ --- by a factor of at least $10^{5}$ to $10^{10}$ at given $\rho $, and depending on $\rho $. We tried also to determine $S(k)$ using an inverse Fourier transform starting from the EM field instead of the EM potential $A_z$: e.g., starting from the component $E_z$ [Eq. (\ref{Ez_mono}) in the time-harmonic case]. This also was not successful in the numerical application with the relevant numbers. Another possibility could be to investigate an approach based on wavelets (e.g. \cite{Ala-et-al2011}). We did not study the feasibility of such an approach for the present problem.\\

Instead, the method we finally used to determine \Couleur{$S_j$} is by the values 
\be\label{S_nj}
S_{n j} := S_j(k_{n j})\quad (n=0,...,N,\quad j=1,..., N_\omega), 
\ee
where 
\be\label{k_nj}
k_{n j} = -K_j + n\delta _j \quad (n=0,...,N)
\ee
is a regular discretization of the integration interval \Couleur{$[-K_j,+K_j]$} for $k$, corresponding with the frequency \Couleur{$\omega _j$}, Eq. (\ref{psi_monochrom_j}). (We remind that $K_j := \omega _j/c$; moreover, $\delta _j : = 2K_j/N$ is the size of the discretization interval.) Using the so-called ``Simpson $\frac{3}{8}$ composite rule", integrals like the one in Eq. (\ref{psi_monochrom_j}) are approximated by discrete sums, as follows:
\be\label{discrete_integral}
\int_{-K_j} ^{+K_j} f(k) \dd k = \sum _{n=0} ^N a_{n j} f(k_{n j}) + O\left(\frac{1}{N^4}\right),
\ee
where $N$ must be a multiple of $3$, and
\bea\label{a_nj}
a_{n j} & = & (3/8)\,\delta _j \quad (n = 0\ \mathrm{or}\ n = N),\\
a_{n j} & = & 2\times (3/8)\,\delta _j \quad (\mathrm{mod}(n,3)=0 \ \mathrm{and}\ n \ne 0\ \mathrm{and}\ n \ne N),\\
a_{n j} & = &  3\times (3/8)\,\delta _j\quad \mathrm{otherwise}.
\eea
Using the approximation (\ref{discrete_integral}) to calculate $\Psi$ [Eqs. (\ref{psi_avec_spectre_2}) and (\ref{psi_monochrom_j})], we get: 
\be\label{Psi'}
\Psi (t,\rho ,z) = \sum _{n=0} ^N \sum _{j = 1} ^{N_\omega } f_{n j}(t,\rho ,z) \,S_{n j} + O\left(\frac{1}{N^4}\right),
\ee
with 
\be\label{f_nj}
f_{n j}(t,\rho ,z) = \frac{\omega _j}{\omega _0} a_n \,J_0\left(\rho \frac{\omega _j}{\omega _0} \sqrt{K_0^2 - k_n^2} \right) \exp \left[ \iC \left( \frac{\omega _j}{\omega _0} k_n z -\omega _j\, t\right ) \right ],
\ee
where the real numbers $a_n \geq 0$ and $k_n$ ($0 \leq n \leq N$) are as $a_{n j}$ and $k_{n j}$ in Eqs. (\ref{a_nj}) and (\ref{k_nj}), replacing $K_j$ by $K_0 =\frac{\omega _0}{c}$, so that
\be\nonumber
a_{n j} = \frac{\omega _j}{\omega _0} a_n, \quad k_{n j} = \frac{\omega _j}{\omega _0} k_n.
\ee
To determine the spectra $S_j$, i.e. the unknown complex numbers (\ref{S_nj}), we fit the sum of the scalar potentials of the individual ``stars", Eq. (\ref{sum_psi_spher_spectre}), by Eq. (\ref{Psi'}). To do that, we evaluate the sum (\ref{sum_psi_spher_spectre}) at a discrete set $G$ of values of $t,\ \rho $ and $z$, that makes a regular three-dimensional grid of points of spacetime:
\be\label{Grid}
G = \{(t_l,\rho _m, z_p),\ 1\leq l\leq N_t,\ 1\leq m\leq N_\rho , \ 1\leq p \leq N_z \}. 
\ee
We group the indices $l, m, p$ as a single index $J = J(l, m, p) \ (1\leq J\leq J_\mathrm{max}= N_t \times N_\rho \times N_z)$. We denote the corresponding values of the sum (\ref{sum_psi_spher_spectre}) by $D_J$:
\be\label{D_J}
D_J = \sum_{i=1} ^{i_\mathrm{max}} \sum_{j=1} ^{N_\omega} \,S'_j\, \frac{\exp(\iC (K_j r_{i m p} -\omega_j t_l))} {K_j r_{i m p}}, \qquad r_{i m p}:= \abs{{\bf x}(\rho _m,\phi=0,z_p) - {\bf x}_i}.
\ee
(Recall that ${\bf x}(\rho,\phi,z )$ is the spatial point with cylindrical coordinates $(\rho , \phi,z )$.) Similarly, we denote $A_{J\ nj} = f_{nj}(t_l,\rho _m, z_p)$. The fitting of the sum (\ref{sum_psi_spher_spectre}) by Eq. (\ref{Psi'}) on the spatiotemporal grid $G$ amounts to solving the linear system
\be\label{LeastSquaresSystem_for_S_nj}
\sum _{n=0} ^N \sum _{j = 1} ^{N_\omega } \, A_{J\ nj} \,S_{n j} = D_J\quad (J=1,...J_\mathrm{max})
\ee
in the sense of the least squares, hence getting the complex numbers $S_{n j}$ as the output. These calculations are implemented on a PC, using the Matlab language and software.




  \subsection{Quadruple precision is needed}\label{Quadruple}

The ratio in Eq. (\ref{kpc/lambda}), thus a number of the order of \ $10^{25}$, gives the magnitude of the spatial argument \ $K_j r_i$\  in the complex exponential \ $e^{\iC \theta}$ \ of Eq. (\ref{sum_psi_spher_spectre}) and the argument of the Bessel function \ $J_0$ \ in Eq. (\ref{psi_monochrom_j}). However, the Bessel \ $J_0$\  function, as well as the real and imaginary parts of \ $e^{\iC \theta}$, oscillate around \ $0$\  with a pseudo-period which is of the order of unity (exactly \ $2 \pi$, for  \ $\cos \theta$  and  \ $\sin\theta$\,). So already to get only the correct {\it sign,} one needs to know their arguments to a precision better than \ $O(1)$. In view of the magnitude of the arguments: \ $O(10^{25})$, it means that\  $25$ \ significant digits are needed to know just the sign of \ $\cos K_j r_i$\  and \ $\sin K_j r_i$\  in Eq. (\ref{sum_psi_spher_spectre}) and the sign of\  $J_0\left(\rho \sqrt{\frac{\omega_j ^2}{c^2}-k^2}\right )$\  in Eq. (\ref{psi_monochrom_j}). Therefore, double precision (16 significant digits) is not enough: quadruple precision (32 significant digits) is needed --- and even, it is not a luxury. Implementing quadruple precision, using the Matlab function $\verb+vpa+$ (for ``variable precision arithmetic"), increases drastically the computation time. But, fortunately, we could reduce significantly the computation time (by a factor of approximately \ $20$ \ for our programs), by using the external toolbox ``Multiprecision Computing Toolbox for Matlab", of Advanpix. In that toolbox, we imposed the number of digits to be \ $41$, in order to reach the same level of numerical precision as with the Matlab function \,$\verb+vpa+$ with default precision, i.e., \ $32$ \ digits plus \ $9$ \ ``guard digits". 
\footnote{\ 
This was advised to the author by Pavel Holoborodko, of Advanpix.
}

\subsection{Calculation of the EM field and its exact character} \label{EM_Field_from_S}

As with Eq. (\ref{psi_avec_spectre_2}) and (\ref{psi_monochrom_j}) for the $A_z$ potential: each of $B_\phi$, $E_\rho$, and $E_z$ is the sum of the time-harmonic components given by Eqs. (\ref{Bphi_mono})--(\ref{Ez_mono}). And as we did with $A_z$ to obtain Eqs. (\ref{Psi'}) and (\ref{f_nj}), we use the approximation (\ref{discrete_integral}) to calculate the integrals in Eqs. (\ref{Bphi_mono})--(\ref{Ez_mono}). We thus get:
\be\label{Bphi'}
B_\phi (t,\rho ,z) = \sum _{n=0} ^N \sum _{j = 1} ^{N_\omega } R_n \,J_1\left(\rho \frac{\omega _j}{\omega _0} R_n \right) \,{\mathcal Re} \left[ F_{n j}(t,z)\right ] + O\left(\frac{1}{N^4}\right),
\ee
\be\label{Erho'}
E_\rho (t,\rho ,z) = \sum _{n=0} ^N \sum _{j = 1} ^{N_\omega } \frac{c^2}{\omega_0 } k_n R_n\,J_1\left(\rho \frac{\omega _j}{\omega _0} R_n \right) \,{\mathcal Re} \left[F_{n j}(t,z)\right ] + O\left(\frac{1}{N^4}\right) ,
\ee
\be\label{Ez'}
E_z (t,\rho ,z) = \sum _{n=0} ^N \sum _{j = 1} ^{N_\omega } \left(\frac{c^2}{\omega_0 } k_n^2 -\omega _0 \right ) \,J_0\left(\rho \frac{\omega _j}{\omega _0} R_n \right) \,{\mathcal Im} \left[F_{n j}(t,z) \right] + O\left(\frac{1}{N^4}\right),
\ee
with \ $R_n = \sqrt{K_0^2 - k_n^2}$ \ and
\be\label{F_nj}
F_{n j}(t,z) = \left(\frac{\omega _j}{\omega _0}\right)^2\, a_n \exp \left[ \iC \left( \frac{\omega _j}{\omega _0} k_n z -\omega _j\, t\right ) \right ]\,S_{n j}.
\ee

\vspace{2mm}
Suppose one starts from an exact solution, with a finite frequency spectrum $(\omega_j)$, of the scalar wave equation: $A_z=\Psi _{(\omega _j)\ (S_j)}$ given by Eqs. (\ref{psi_avec_spectre_2}) and (\ref{psi_monochrom_j}) with exact ``wave-vector spectra'' $S_j$. Then Eqs. (\ref{Bphi'})--(\ref{Ez'}) for the EM field give in general only an approximation of the associated EM field (defined by summing the contributions (\ref{Bphi_mono})--(\ref{Ez_mono})): this is due to the discrete integration method (\ref{discrete_integral}), using the discrete values (\ref{S_nj}) of the exact functions $S_j$. \\

 However, the integration formula (\ref{discrete_integral}) is {\it exact} (i.e., the remainder is not only $O(1/N^4)$ but exactly zero) if the integrand function $f$ is a polynomial of degree $\le 3$. This is because the remainder is proportional to $f^{(4)}(\xi)$ for some $\xi \in ]-K_j,+K_j[$ \ \cite{Atkinson1989}. Moreover, given e.g. the first four arguments $k_{n j}$ and the four corresponding function values $S_{n j}$ ($n=0,...,3$), there exists one and only one polynomial $P$ of degree $\le 3$, such that $P(k_{n j})=S_{n j}$ ($n=0,...,3$). (This is the well-known ``Unisolvence theorem''; note that we are considering a fixed value of the frequency index $j$.) By construction, $N$ is a multiple of $3$, hence the whole integration interval $[-K_j,+K_j]$ is the union of $N/3$ adjacent subintervals, each covering three steps $\delta_j$. Thus, due to the unisolvence theorem: in each of those subintervals, the four successive arguments $k_{n j}$ and the four corresponding function values $S_{n j}$ define a unique 3rd-degree polynomial, for which the integration (\ref{discrete_integral}) is exact. It follows that the integration (\ref{discrete_integral})  is exact for the piecewise polynomial function $\widetilde{S}_j$ which continuously extends those $N/3$ polynomial functions to the whole interval $[-K_j,+K_j]$. 
\footnote{\ 
The continuity of $\widetilde{S}_j$ results from the common value $S_{n j}$ at the common bound of any two successive subintervals. (The derivatives of $\widetilde{S}_j$ are in general not continuous at the bounds of the subintervals, though.) Note that the integration formula (\ref{discrete_integral}) gives the same result whether it is applied to the whole interval, or successively to each of the subintervals, each covering three steps $\delta_j$, because the weights $a_{n j}=(3/8)\,\delta _j$ at the bounds of two successive subintervals add to give the weight $a_{n j}=2\times (3/8)\,\delta _j$ at a multiple of three steps inside the whole interval.
}
Therefore, Eqs. (\ref{Psi'}) and (\ref{f_nj}) actually define an exact solution $\widetilde{\Psi}$ of the scalar wave equation, which corresponds with substituting the functions  $\widetilde{S}_j$ for $S_j$ in Eqs. (\ref{psi_avec_spectre_2}) and (\ref{psi_monochrom_j}). Similarly, Eqs. (\ref{Bphi'})--(\ref{Ez'}) for the EM field give the exact result of adding the contributions (\ref{Bphi_mono})--(\ref{Ez_mono}) for the different frequencies $\omega_j$, when in these contributions one considers, for the frequency $\omega=\omega_j$, the spectrum function $S=\widetilde{S}_j$. In other words, {\it Eqs. (\ref{Bphi'})--(\ref{Ez'}) provide an exact solution of the source-free Maxwell equations, deduced from an exact solution  (\ref{Psi'}) of the scalar wave equation} --- all corresponding with the spectrum functions $\widetilde{S}_j$, which are piecewise 3rd-degree polynomials.

\subsection{Validation test}\label{Correctness}

The formulas (\ref{Psi'}) and (\ref{Bphi'})--(\ref{Ez'}) were implemented numerically and that numerical implementation was tested for the case with spherical symmetry, as follows. We can define an exact solution of the source-free Maxwell equations by Eqs. (\ref{Bfi})--(\ref{Ez}), with $A_z = \psi _{\omega\ S \equiv  \frac{c}{2\omega }}$ the spherically-symmetric time-harmonic solution (\ref{psi_spher}) of the scalar wave equation. This yields (after taking the real part):
\be\label{Bphi_spher}
B_\phi (t,\rho ,z) = \frac{\rho \cos \omega t}{r^2}\,\left(\frac{\sin Kr}{Kr}-\cos Kr \right),
\ee
\be\label{Erho_spher}
E_\rho (t,\rho ,z) = \frac{c^2}{\omega} \frac{\rho z}{r^3} \sin \omega t \left [ -\frac{3\cos Kr}{r}  + \left(\frac{3}{K r^2} - K\right) \sin Kr \right ],
\ee
\be\label{Ez_spher}
E_z (t,\rho ,z) = \frac{c^2}{\omega} \frac{\sin \omega t}{r} \left [ \left (1-\frac{3z^2}{r^2}\right ) \frac{\cos Kr}{r}   + \left (\frac{3z^2}{Kr^3}-\frac{1+K^2 z^2}{Kr}+Kr\right ) \frac{\sin Kr}{r} \right],
\ee
with $K:=\omega/c$ and $r:= \sqrt{\rho ^2+z^2}$. The outputs of the (exact) Eqs. (\ref{Bphi_spher})--(\ref{Ez_spher}) were compared to those of Eqs. (\ref{Bphi'})--(\ref{Ez'}) applied with the relevant constant spectrum $S(k):=\frac{c}{2\omega }$, thus in Eq. (\ref{F_nj}):
\be\label{Sn_spher}
S_n := S(k_n) = \frac{c}{2\omega } \quad (n=0,...,N).
\ee
(We are considering the single angular frequency case: $N_\omega=1$, hence the index $j$ is omitted since it takes only one value: $j=1$; moreover, we set $\omega_1:=\omega_0:=\omega$ in Eqs. (\ref{Bphi'})--(\ref{F_nj}).) Note that, with the exact spectrum $S(k):=\frac{c}{2\omega }$, Eqs. (\ref{Bphi_mono})--(\ref{Ez_mono}) provide {\it just the same} exact fields as do Eqs. (\ref{Bphi_spher})--(\ref{Ez_spher}). However, even with the exact spectrum values (\ref{Sn_spher}), Eqs. (\ref{Bphi'})--(\ref{Ez'}) provide only an approximation of {\it those} exact fields, due to the discrete integration (\ref{discrete_integral}). \\

Figs. 1 to 3 show, for the three different scales investigated, the relative average quadratic differences between the fields $\Psi = A_z$, $B_\phi$, $E_\rho$, $E_z$, as calculated either ``directly'', i.e., by Eqs. (\ref{psi_spher}) and (\ref{Bphi_spher})--(\ref{Ez_spher}), or ``with the spectrum (\ref{Sn_spher})'', i.e., by Eqs. (\ref{Psi'}) and (\ref{Bphi'})--(\ref{Ez'}) applied with the spectrum values (\ref{Sn_spher}). The different scales were here: \verb+scale+ = $10^n \lambda$ ($n=1,2,3$), with $\lambda : = c/\nu$. For this test, the frequency was taken to be $\nu:=\omega/(2 \pi) =100\, \mathrm{MHz}$, thus $\lambda=3\,\mathrm{m}$. The average quadratic differences are in general evaluated on a regular three-dimensional spatio-temporal grid (\ref{Grid}) for the variables $t, \rho, z$. Thus $t$ takes $N_t$ values between $t_0$ and $t_0+(N_t-1)\delta t$, $\rho$ takes $N_\rho$ values between $\rho_0$ and $\rho_0+(N_\rho-1)\delta \rho$, and $z$ takes $N_z$ values between $z_0$ and $z_0+(N_z-1)\delta z$, with $\delta t=T/N_t$, $\delta \rho=\verb+scale+/N_\rho$, and $\delta z= \verb+scale+/(10N_z)$. However, in the present case with harmonic time dependence, we took $N_t=1$, thus one value of $t=t_0$, which was fixed at $T/8$, with $T=1/\nu=10^{-8}\,\mathrm{s}$. So here the grid is two-dimensional in fact. Also, we took here $N_\rho=14$, $N_z=13$, and $\rho_0=\delta \rho$,
 $z_0=5 \delta z$.\\

\begin{figure}[ht]
\centerline{\includegraphics[height=9cm]{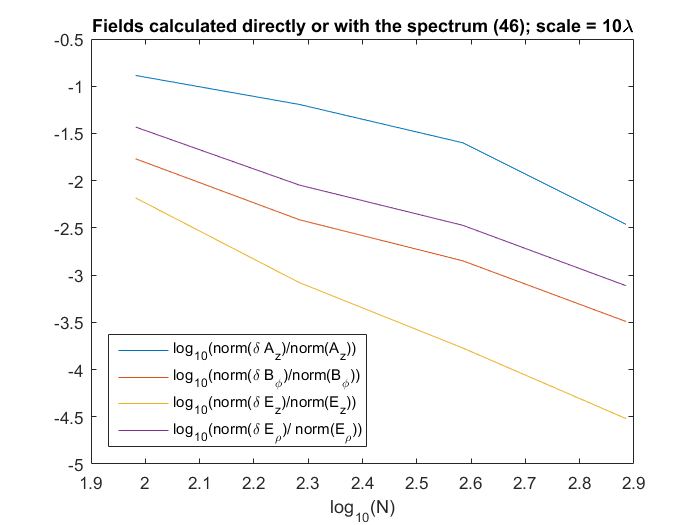}}
\caption{Average quadratic differences on a $(\rho,z)$ grid. Scale: $10\lambda$}
\end{figure}
\begin{figure}[ht]
\centerline{\includegraphics[height=9cm]{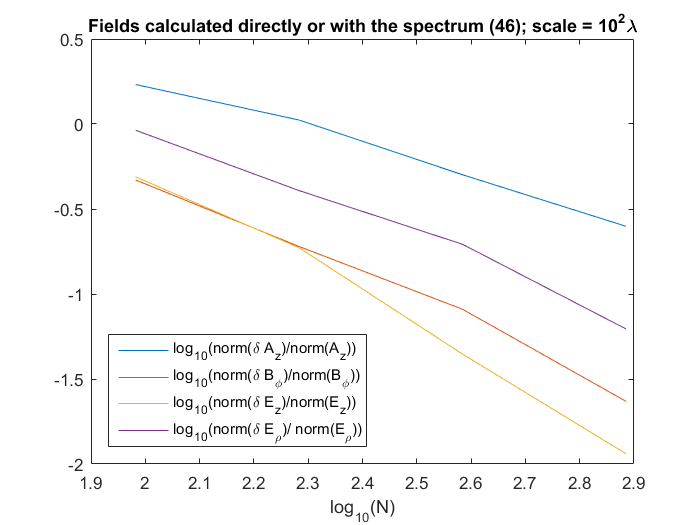}}
\caption{Average quadratic differences on a $(\rho,z)$ grid. Scale: $10^2\lambda$}
\end{figure}
\begin{figure}[ht]
\centerline{\includegraphics[height=9cm]{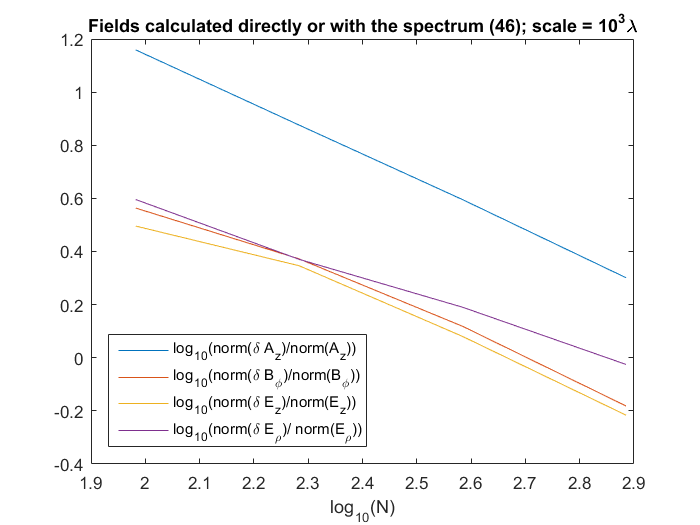}}
\caption{Average quadratic differences on a $(\rho,z)$ grid. Scale: $10^3\lambda$}
\end{figure}

\hypertarget{error&scale}{As expected,} the errors (the relative delta's on the different fields calculated either ``directly'' or ``with the spectrum (\ref{Sn_spher})'') decrease strongly as the discretization of the (here constant) spectrum function $S(k)$ becomes finer, i.e., with increasing $N$. (See Eq. (\ref{S_nj}).) {\it This validates the correctness of our calculations.} However, the errors increase quickly when the scale length $\verb+scale+$ is increased (even though it remains here enormously smaller than the galactic scale). This is because the integrals in Eqs. (\ref{psi_monochrom_j}) and (\ref{Bphi_mono})--(\ref{Ez_mono}) involve functions of $k$ that oscillate with a frequency or a pseudo-frequency which is proportional to the magnitude of the spatial variables $\rho$ and $z$. As these integrals are approximated by the discrete sums (\ref{Psi'}) and (\ref{Bphi'})--(\ref{Ez'}), one would have to increase the discretization number \Couleur{$N$} in proportion of the scale length $\verb+scale+$.\\

In addition to this validation test, which does test our programs but is based on the ``unphysical" solution (\ref{psi_spher}), we will also compare the outcome of Eqs. (\ref{Bphi'})--(\ref{Ez'}) with the exact solution that corresponds to a set of spherical sources situated at the points ${\bf x}_i$. That is, one defines an exact solution of the source-free Maxwell equations by summing solutions obtained by Eqs. (\ref{Bfi})--(\ref{Ez}) applied to the outgoing spherical wave (\ref{psi_spher'}). That sum is similar to the sum $\Psi _{({\bf x}_i)\ (\omega _j)\ (S'_j)}$ given by Eq. (\ref{sum_psi_spher_spectre}). (Thus, all sources have the same amplitude and the same frequency spectrum, but this can easily be changed.) The fields produced by the individual source at ${\bf x}_i$ are denoted by $B_{\phi\,i}$, $E_{\rho\,i}$ and $E_{z\,i}$. In the case of a single frequency $\omega $, one first applies Eqs. (\ref{Bfi})--(\ref{Ez}) with $A_z=\psi' _\omega $ as given by Eq. (\ref{psi_spher'}):
\be\label{Bphi_spher'}
B_{\phi \omega } (t,\rho ,z) = \frac{\rho}{r^2} \left(\sin \varphi +\frac{\cos \varphi }{Kr} \right),\qquad \varphi :=Kr - \omega t,
\ee
\be\label{Erho_spher'}
E_{\rho \omega } (t,\rho ,z) = \frac{c^2}{\omega} \frac{\rho z}{r^3} \left [ 3\frac{\cos \varphi }{r}  + \left( K-\frac{3}{K r^2} \right) \sin \varphi  \right ],
\ee
\be\label{Ez_spher'}
E_{z \omega } (t,\rho ,z) = \frac{c^2}{\omega r} \left [ \left (\frac{3z^2}{r^2} - 1 \right ) \frac{\cos \varphi }{r} + \left (-\frac{3z^2}{K r^3}+\frac{1+K^2 z^2}{K r}-Kr \right ) \frac{\sin \varphi }{r} \right],
\ee
but with $\rho$, $z$ and $r$ being replaced by 
\be\label{rho'_i}
\rho'_i := \sqrt{(x-x_i)^2 + (y-y_i)^2},
\ee 
\be\label{Z_i}
Z_i := z-z_i,
\ee 
\be\label{r_i}
r_i :=\abs{{\bf x}-{\bf x}_i} = \sqrt{(x-x_i)^2 + (y-y_i)^2 +(z-z_i)^2} = \sqrt{\rho'^2 _i +Z_i ^2}.
\ee 
The definition of $B_{\phi\,i}$, $E_{\rho\,i}$ and $E_{z\,i}$ is that the exact fields produced at ${\bf x}$ by the source at ${\bf x}_i$  are decomposed on the orthonormal direct basis made of
\be\label{e'_rho}
{\bf e}'_{\rho \, i} = {\bf e}'_\rho({\bf x};{\bf x}_i)= \left ( (x-x_i) {\bf e}_x + (y-y_i){\bf e}_y \right )/\rho'_i,
\ee 
\be\label{e'_phi}
{\bf e}'_{\phi \, i} = {\bf e}'_\phi({\bf x};{\bf x}_i) := {\bf e}_z \wedge {\bf e}'_\rho ({\bf x};{\bf x}_i) = \left ( (x-x_i) {\bf e}_y - (y-y_i){\bf e}_x \right )/\rho'_i,
\ee
and ${\bf e}_z$. When each spherical source has the same finite frequency spectrum  $(\omega_j,S'_j)$, as in Eq. (\ref{sum_psi_spher_spectre}), it is understood that the components $B_{\phi\,i}$, $E_{\rho\,i}$ and $E_{z\,i}$ involve the corresponding weighted sum, e.g.
\be\label{B_phi i}
B_{\phi\,i} = \sum _{j=1} ^{N_\omega}  S'_j \ B_{\phi \,i \, {\omega_j}},
\ee
where $B_{\phi \,i \, {\omega_j}}$ is given by Eq. (\ref{Bphi_spher'}) with $\rho'_i, Z_i, r_i, \omega_j, K_j$ in the place of $\rho, z, r, \omega, K$. The total exact fields, sums of these different fields, are (reminding that $B_{\rho \, i} = B_{z \, i} = E_{\phi \, i} =0$):
\be\label{B_phi tot}
B_\phi := {\bf B.e}_\phi = \sum_i  B_{\phi \, i} \,{\bf e}'_{\phi \, i}{\bf . e}_{\phi},
\ee
\be\label{E_rho tot}
E_\rho := {\bf E.e}_\rho = \sum_i  E_{\rho \, i}\, {\bf e}'_{\rho \, i}{\bf . e}_{\rho},
\ee
\be\label{E_z tot}
E_z := {\bf E.e}_z = \sum_i  E_{z \, i} .
\ee
In general, the other components of the total fields may be non-zero also, but we assume that the distribution of the identical spherical sources is axisymmetric (see Sect \ref{StarDistribution}). In that case, we expect that $E_\phi = B_\rho =B_z =0$. Note also that ${\bf  e}_{\rho}$ and ${\bf  e}_{\phi}$, hence also the scalar products in Eqs. (\ref{B_phi tot})--(\ref{E_rho tot}), and therefore also the $B_\phi$ and $E_\rho$ components, are not defined if $\rho=0$.

\section{Results  and discussion}\label{Results}

\begin{figure}[ht]
\centerline{\includegraphics[height=9cm]{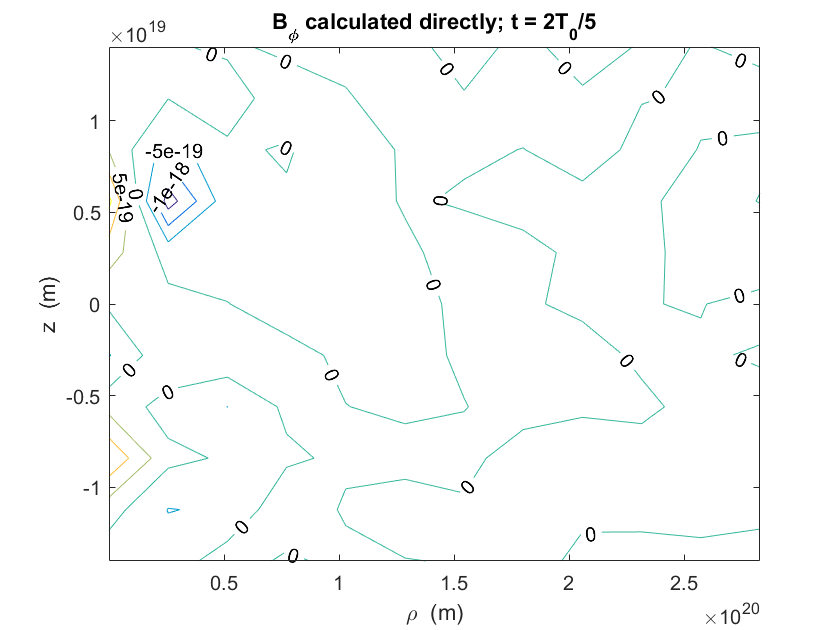}}
\centerline{\includegraphics[height=9cm]{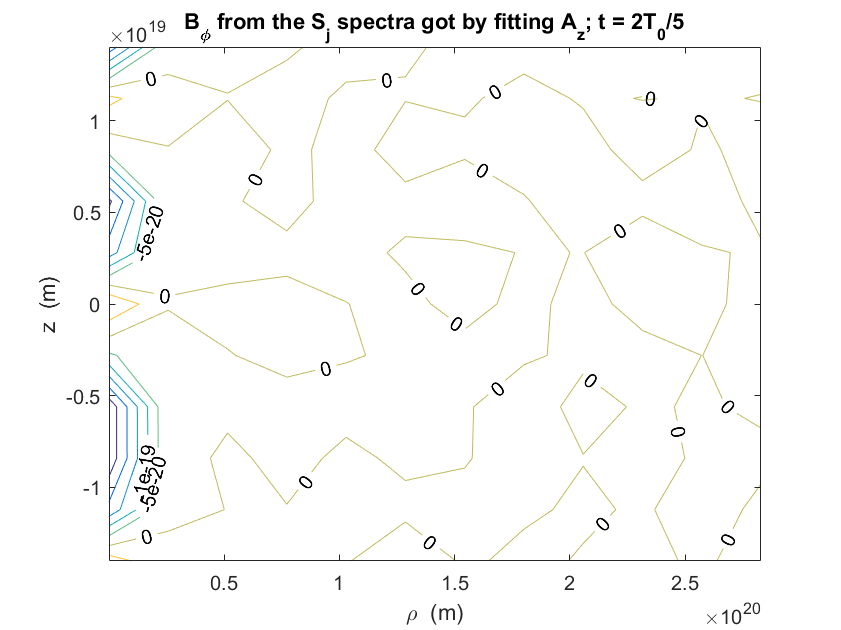}}
\caption{$B_\phi$ calculated directly or from the model; $t=2T_0/5$}
\end{figure}

A file of some $10^4$ randomly generated ``stars" (as described in Subsect. \ref{StarDistribution}) has been used here, more precisely one with $16 \times 16\times 36$ triplets $(\rho ,z, \phi )$. \\

The angular frequencies $\omega _j$ are the same in the expression to be fitted, Eq. (\ref{sum_psi_spher_spectre}), and in the analytical expression used to fit it, Eq. (\ref{psi_avec_spectre_2}). They are regularly spaced and symmetric around a central frequency $\omega _0$, thus
\be\label{omega_j}
\omega _j = \omega _0-\Delta \omega +(j-1)\frac{\Delta \omega }{N_\mathrm{inter}}, \quad (j=1,..., N_\omega = 2N_\mathrm{inter}+1),
\ee
where $\omega _0 = \frac{2\pi\,c}{\lambda _0}$, $\Delta \omega < \omega _0$. In the calculations, we took $N_\mathrm{inter} =5$ (hence $N_\omega =11$), $\lambda _0 := 0.5\,\mu \mathrm{m}$, $\Delta \omega=\omega _0/2$. The weights $S'_j$ affected to the different frequencies in Eq. (\ref{sum_psi_spher_spectre}) have the form
\be
S'_j \propto \omega _j \exp\left[-\frac{1}{2\sigma ^2}\left(\omega _j-\omega _0\right)^2 \right] \quad (j=1,...,N_\omega )
\ee 
with \Couleur{$\sigma = \Delta \omega $}, and are normed so that $\sum_j \,S'_j=1$ .\\

A few different spacetime domains (variables \Couleur{$t,\rho ,z$}) of galactic dimensions have been used. The adopted sizes of the domain for the calculations discussed here were as follows:
\bea\label{Var_t}
0 & \leq & t \ < \ T_0  \ := \ \frac{\lambda _0}{c}, \\
\label{Var_rho}
 \rho_0 & \leq &\rho \ < \verb+scale+\ , \\
 \label{Var_z}
-\verb+scale+/20 & < & z \ < \verb+scale+/20,
\eea
with $\verb+scale+=3.086 \times 10^{20}\, \mathrm{m} \ \simeq \ 10\, \mathrm{kpc}$ and $\rho_0=\verb+scale+/10^{6}$ in these calculations. \\

We have discretized that domain using grids (\ref{Grid}) with \ $N_t = 4,\quad N_\rho = 8, \quad N_z = 7$ --- in short $(4,8,7)$ ---, or $(5,12,11)$, or $(7,14,13)$. The symbols \ $\leq $\ and \ $<$\  in Eqs. (\ref{Var_t})--(\ref{Var_z}) are meant to indicate that, e.g. for Eq. (\ref{Var_t}), the discrete variation of \ $t$ \ begins with \ $t=0$ \ and ends with the largest multiple of the time step that is smaller than $T_0$. The time step is  $\delta t = T_0/N_t$, so \ $t= (i_t-1)\delta t \quad (i_t=1,..., N_t)$ --- and similarly for $\rho$ and $z$ in Eqs. (\ref{Var_rho})--(\ref{Var_z}). (This is just the same kind of variation as for the validation test of Sect. \ref{Correctness}, but here $\verb+scale+$ has a value that is relevant to a galaxy.) Thus, while we browse a very small total interval of time using a very small time step, we do scan a large spatial scale, representative of a disk galaxy. The reason for imposing this difference is that the variation of the fields has a quasi-periodic character in time. This has been checked for the $A_z$ potential by calculating the sum (\ref{sum_psi_spher_spectre}) with a very large time step, close to $\delta \rho/c$. Whereas, as we will see below, the fields have a definite spatial variation. On the other hand, the following values were tried for the number $N$ in Eqs. (\ref{S_nj})--(\ref{Psi'}): $N=6, 12, 24, 48, 96, 192$. \\

We will compare the field components $B_\phi$, $E_\rho$, $E_z$, as calculated either ``directly'', i.e., from Eqs. (\ref{B_phi tot})--(\ref{E_z tot}), or ``from the model'', i.e., from Eqs. (\ref{Bphi'})--(\ref{Ez'}), using in the latter case in Eq. (\ref{F_nj}) the spectrum values $S_{n j}$ obtained from  fitting the sum (\ref{sum_psi_spher_spectre}) by Eq. (\ref{Psi'}), Eq. (\ref{LeastSquaresSystem_for_S_nj}). Thus, in both cases, the field derives from exact solutions of the scalar wave equation by Eqs. (\ref{A_from_Az}) and (\ref{Bfi})--(\ref{Ez}): either the spherical functions entering the sum (\ref{sum_psi_spher_spectre}), or the function (\ref{Psi'}), which is obtained precisely from fitting the sum (\ref{sum_psi_spher_spectre}). \\

Figures 4 to 9 show, for the $(5,12,11)$ spatiotemporal grid, the contour levels of the field components $B_\phi$, $E_\rho$, $E_z$, as calculated either ``directly'' or ``from the model'' --- in the latter case, on the same spatiotemporal grid $(5,12,11)$ used for the fitting, and with $N=48$. In order to save place, we selected somewhat arbitrarily, and independently for each component, $3$ values of the time among the available $5$ values. Also, very similar figures are obtained if one uses another spatiotemporal grid, like $(4,8,7)$ or $(7,14,13)$. \\

\begin{figure}[ht]
\centerline{\includegraphics[height=9cm]{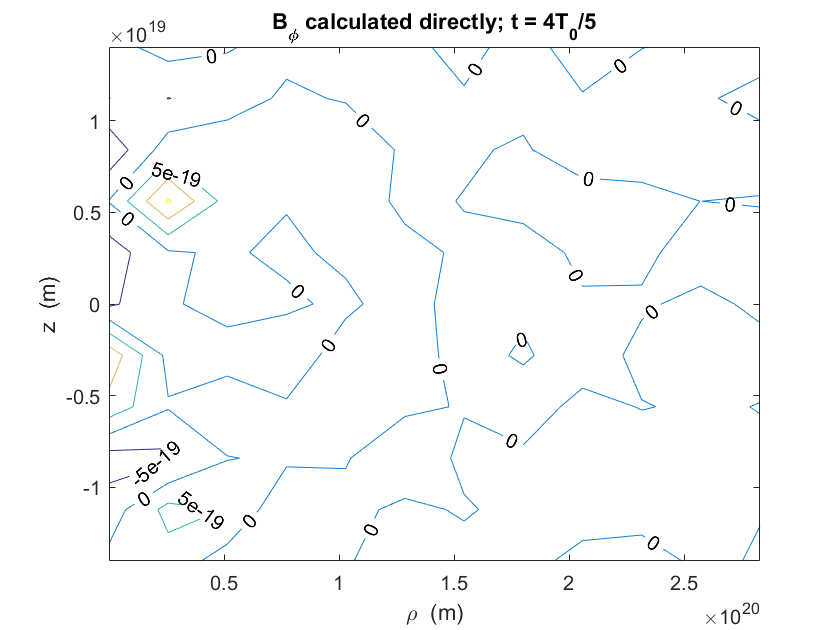}}
\centerline{\includegraphics[height=9cm]{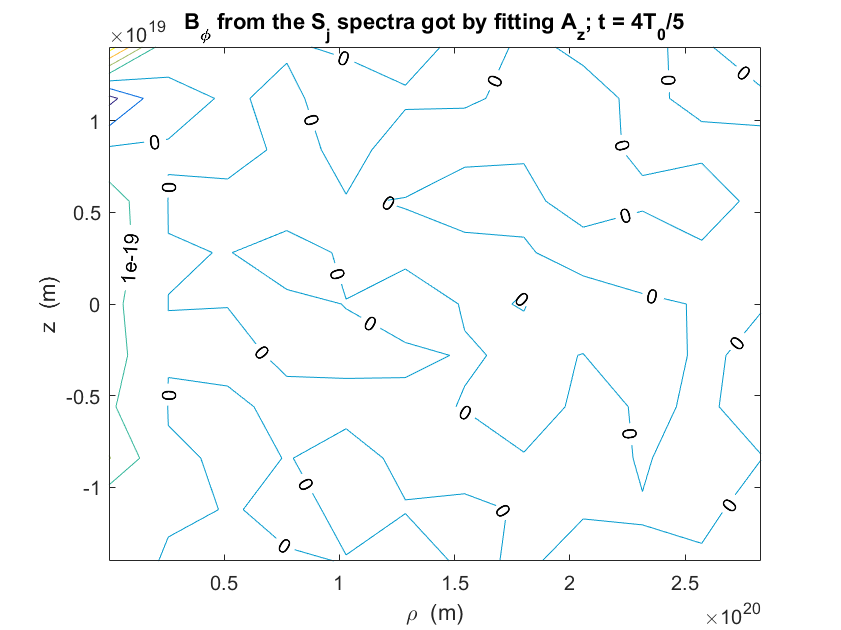}}
\caption{$B_\phi$ calculated directly or from the model; $t=4T_0/5$}
\end{figure}

\begin{figure}[ht]
\centerline{\includegraphics[height=9cm]{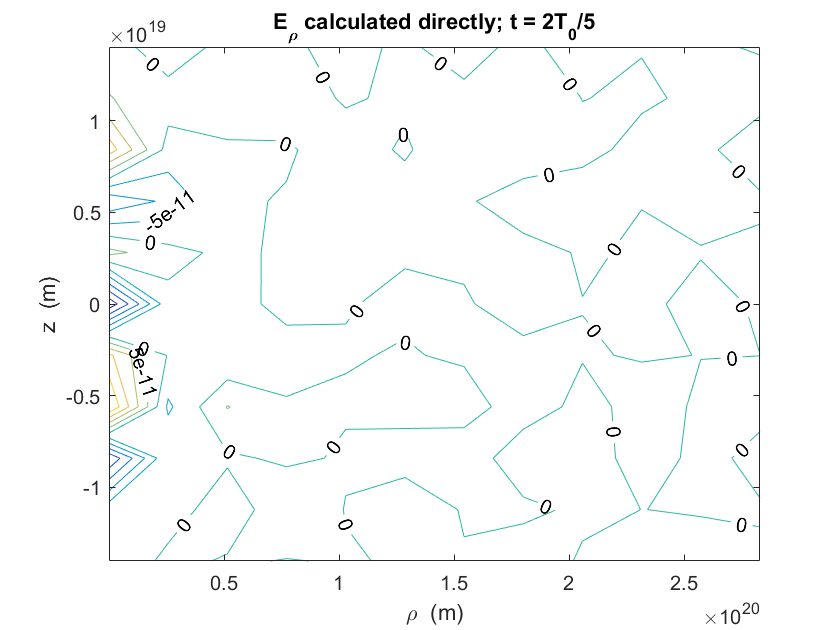}}
\centerline{\includegraphics[height=9cm]{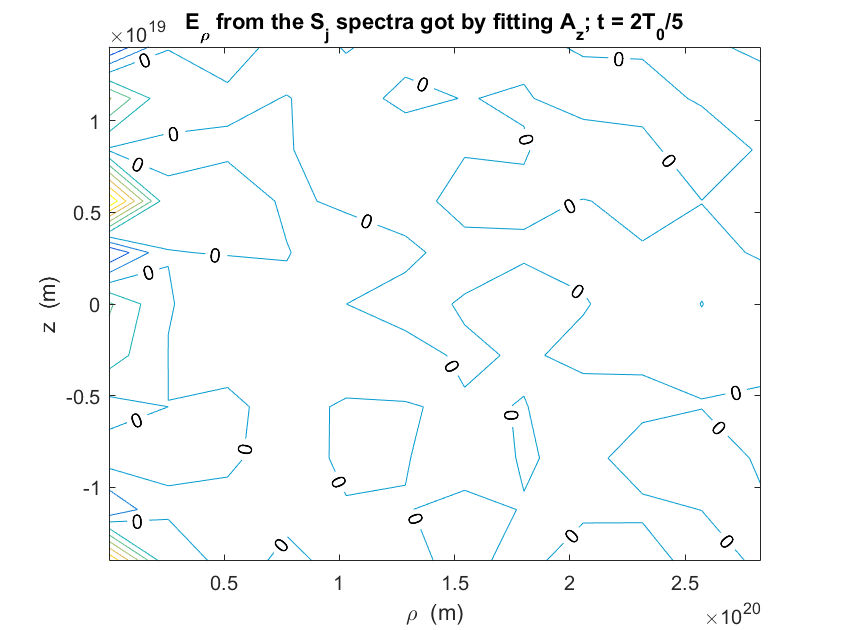}}
\caption{$E_\rho$ calculated directly or from the model; $t=2T_0/5$}
\end{figure}

\begin{figure}[ht]
\centerline{\includegraphics[height=9cm]{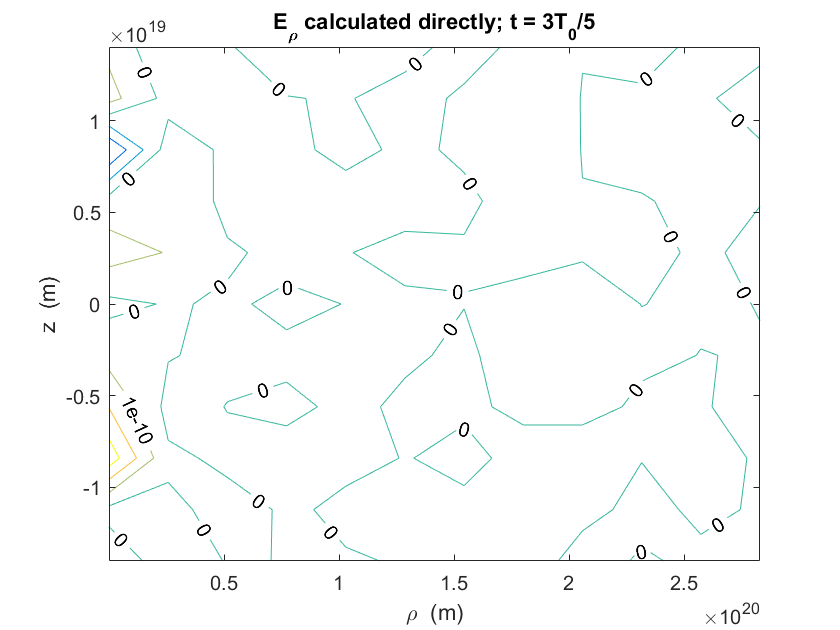}}
\centerline{\includegraphics[height=9cm]{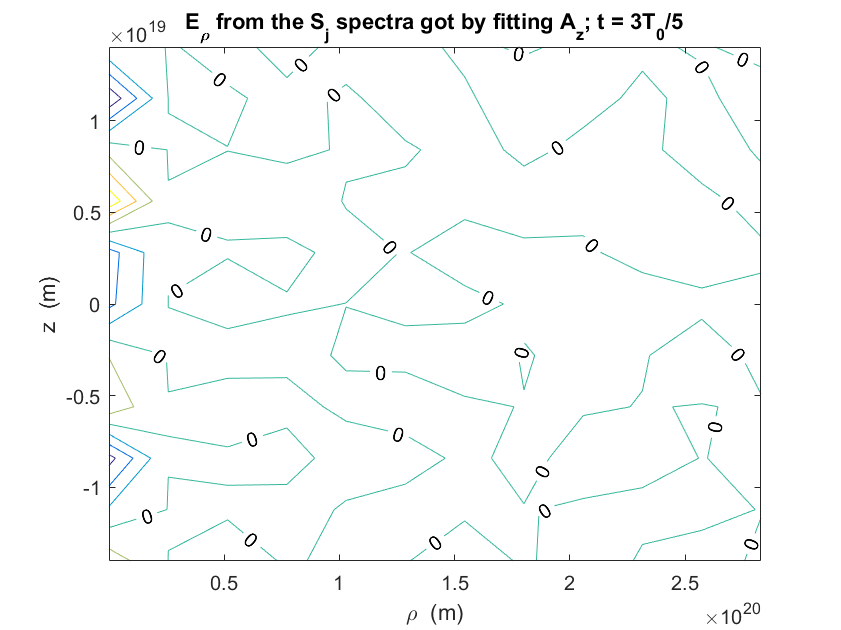}}
\caption{$E_\rho$ calculated directly or from the model; $t=3T_0/5$}
\end{figure}

\begin{figure}[ht]
\centerline{\includegraphics[height=9cm]{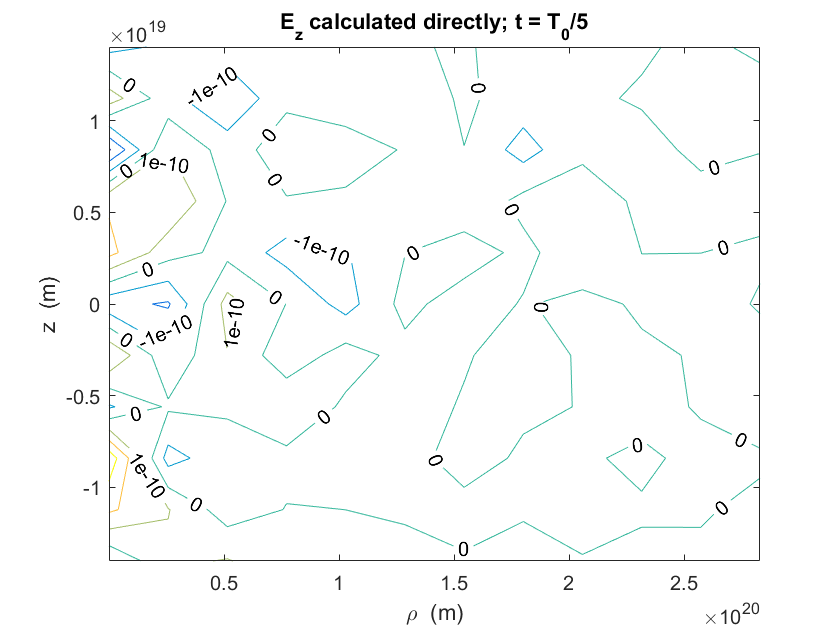}}
\centerline{\includegraphics[height=9cm]{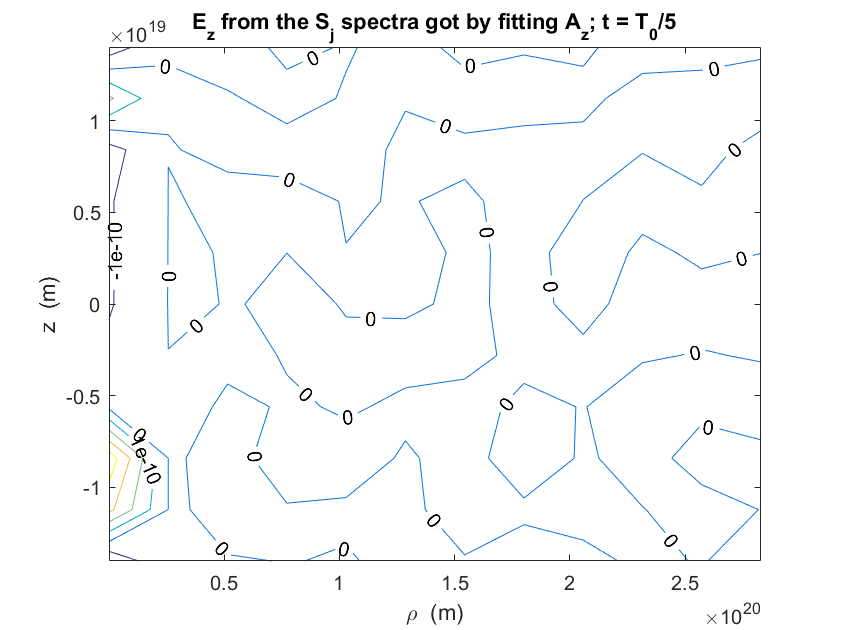}}
\caption{$E_z$ calculated directly or from the model; $t=T_0/5$}
\end{figure}

\begin{figure}[ht]
\centerline{\includegraphics[height=9cm]{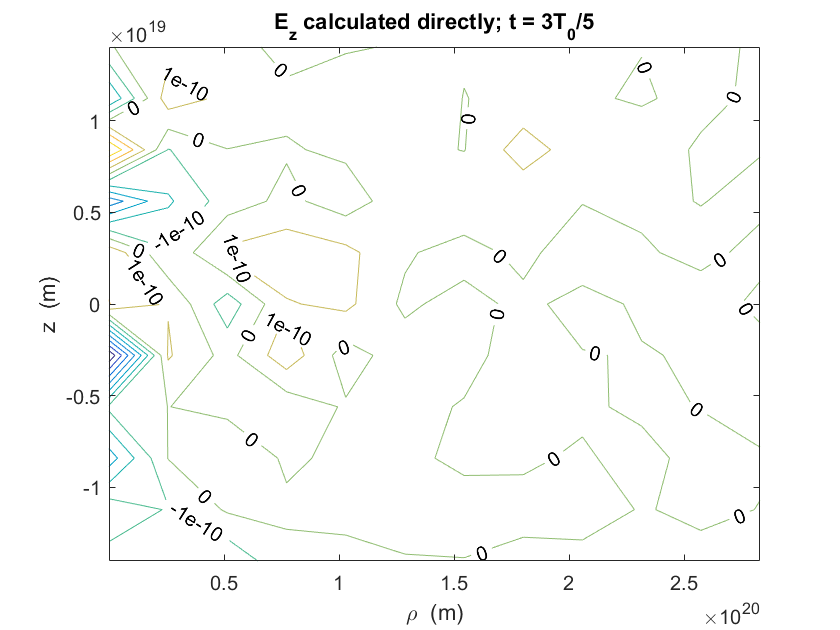}}
\centerline{\includegraphics[height=9cm]{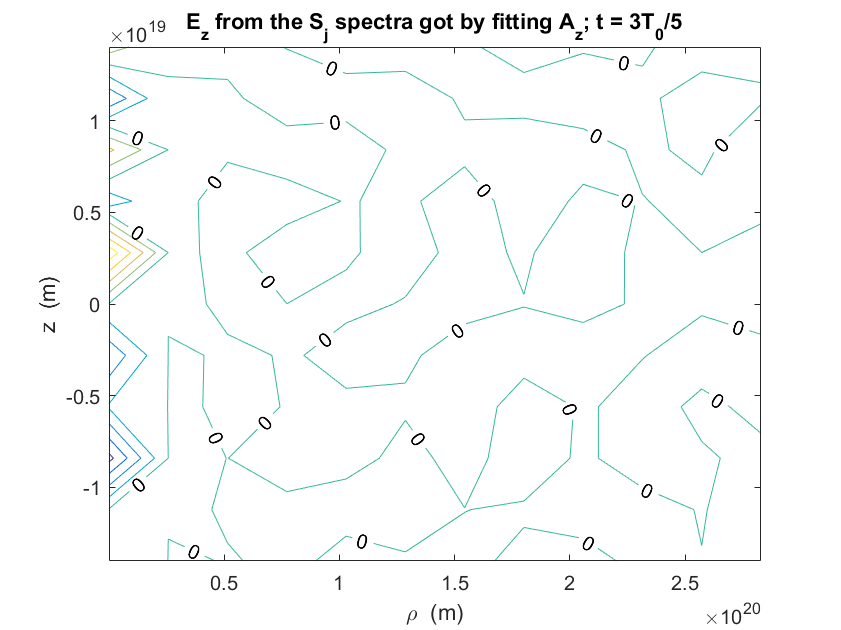}}
\caption{$E_z$ calculated directly or from the model; $t=3T_0/5$}
\end{figure}

If the \hyperlink{error&scale}{increase of the error with scale}, found for the smaller scales investigated in the validation test of Subsect. \ref{Correctness}  (up to $10^3 \lambda$), would continue up to the scale relevant to a typical disc galaxy ($\simeq 3\times 10^{25}\lambda$), then the relative quadratic errors between the field components calculated either directly or from the model (e.g. $\parallel  \delta B_\phi \parallel /\parallel B_\phi \parallel$) would reach huge values for the galactic scale that we are investigating in this section. For the  \hyperref[Correctness]{validation test},  the increase of the error with scale was due to the fact that the discretization number \Couleur{$N$} was not increased in proportion of the scale length. The exact spectrum (\ref{exact spectrum spherical}) was available, and the discretized spectrum values were taken from it, Eq. (\ref{Sn_spher}). In contrast, in this section, the discretized spectrum values $S_{n j}$  are now obtained from fitting the sum (\ref{sum_psi_spher_spectre}) by Eq. (\ref{Psi'}) over a spatio-temporal grid with galactic-like spatial dimensions, Eqs. (\ref{Var_rho})-(\ref{Var_z}). For the present calculation, the relative quadratic errors are rather close to unity. The qualitative features of the fields are the same for the fields calculated directly or with the model, i.e., from the spectra obtained by fitting the $A_z$ potential:\\

i) The fields are more intense close to the $z$ axis. This feature is true for both the direct calculation and the model, but it appears more clearly with the model. Moreover, using that model,  based on Eqs. (\ref{Bphi'})--(\ref{Ez'}), one can calculate the three components $B_\phi$, $E_\rho$ and $E_z$ also for $\rho=0$ --- which is not the case for the $B_\phi$ and $E_\rho$ components when one uses the direct calculation based on Eqs. (\ref{B_phi tot})--(\ref{E_z tot}), see after those equations. The model gives $B_\phi=E_\rho = 0$ for $\rho=0$, because $J_1(0)=0$. In contrast,  for $E_z$, that model predicts very high values for $\rho=0$, of the order $E_z=O(10)$ with the spectrum values obtained by fitting the sum (\ref{sum_psi_spher_spectre}) by Eq. (\ref{Psi'}) on the spacetime domain (\ref{Var_t})--(\ref{Var_z}),  using e.g. the parameters described in the paragraph following Eqs. (\ref{Var_t})--(\ref{Var_z}) --- but using these spectrum values to calculate the fields on a shifted grid, with $\rho $ starting at $\rho_0=0$ instead of $\rho_0=\verb+scale+/10^{6}$. \\

ii) The maximum intensities (positive and negative), calculated either directly or from the model, have quite similar values --- but the positions of the maxima are generally different between the direct calculation and the model, except for the fact mentioned, that they are close to the $z$ axis. \\

An important point has to be noted in this connection. As we saw, the fields have a definite spatial variation at the galactic scale, in contrast to their quasi-periodic time variation with a very small time period $T_0 = \lambda_0/c$. However, to that large-scale spatial variation, is superimposed an oscillatory variation at the very small scale of the main wavelength $\lambda_0=0.5\,\mu$m. This results again from the analytical expressions of the fields, e.g. it is easy to see in Eqs. (\ref{Bphi'})--(\ref{Ez'}) for the model. 
\footnote{\
However, the same remark applies to the direct calculation, e.g. Eq. (\ref{B_phi tot}) for $B_\phi$, with $B_{\phi \, i}$ given by Eq. (\ref{B_phi i}), where $B_{\phi \,i \, {\omega_j}}$ is given by Eq. (\ref{Bphi_spher}) with $\rho'_i, Z_i, r_i, \omega_j, K_j$ in the place of $\rho, z, r, \omega, K$.
 }
The oscillatory spatial variation at the wavelength scale implies a high sensitivity of the details of the calculations on a big spatiotemporal grid like (\ref{Var_t})--(\ref{Var_z}) to small variations of the parameters. This is seen, for example, when the spectrum functions $S_j$ obtained from a fitting are used to calculate the fields on a different spatiotemporal grid than the one used for the fitting\,; or, when two different orders $N$ are used for the same grid. However, the main features of the calculations, as described at points i) and ii) above, are robust, and the relative quadratic differences between different calculations on the same grid usually remain of the order of unity. An exception is if a too low order $N$ is used (e.g. $N=6$ used to fit and calculate with the model on the $(5,12,11)$ grid), in which case larger differences can exist.
\\

Remind that the schematization which leads to the direct calculation (\ref{B_phi tot})--(\ref{E_z tot}) is a rather simple one: an axisymmetric disk-like distribution of point sources, each of which emitting an EM field deriving via Eq. (\ref{A_from_Az}) from a spherically symmetric solution $\Psi$ of the scalar wave equation.  As we saw at the end of Subsect. \ref{EM_Field_from_S}, also the field got from the model is an exact axisymmetric solution of the source-free Maxwell equations. The set of point-like sources, that may represent a disk galaxy, and that leads directly to the calculation (\ref{B_phi tot})--(\ref{E_z tot}), is also used to adjust the model. The latter, however, is based on the nonsingular (``continuous'') equations  (\ref{Psi'}) and (\ref{Bphi'})--(\ref{Ez'}) --- in contrast with Eqs. (\ref{sum_psi_spher_spectre}) and  (\ref{B_phi tot})--(\ref{E_z tot}), that are singular at each source (i.e., for ${\bf x} ={\bf x}_i$). Therefore, the field obtained from the model is at least as representative of the EM radiation field in a disc galaxy as the field calculated directly can be. \\

  


\section{Conclusion}\label{Conclusion}

In this work, an analytical model has been built for the Maxwell field in an axisymmetric galaxy, in particular for that field which results from stellar radiation. This model is based on a representation of any totally propagating axisymmetric source-free Maxwell field as the sum of two fields given explicitly: in the case of a time-harmonic field, the first field is given by Eqs. (\ref{Bphi_mono})--(\ref{Ez_mono}), and the second one is deduced by the duality (\ref{dual}) from a field of the same form. In a previous work, the general applicability of this representation has been \hyperref[Theorem]{proved}. \\

The model is adjusted by fitting to it the sum of spherical radiations emitted by a set of point-like ``stars". The distribution of these point-like objects is axisymmetric. It builds a flat disk, symmetrical with respect to a plane perpendicular to the symmetry axis, and the dimensions have been chosen to represent a disk galaxy similar to the Milky Way. The model provides an exact solution of the source-free Maxwell equations, also after the discretization that is used to calculate the relevant integrals. \\

The huge ratio distance/wavelength needs to implement a numerical precision better than the quadruple precision. The model and the corresponding software have passed a validation test based on an exact solution with spherical symmetry. The results for a disk galaxy indicate that the field is highest near to the $z$ axis, and there the $E_z$ component dominates over $E_\rho$. In a further stage, it will be possible to adjust the model so as to very accurately describe the measured local EM spectrum. The model presented will then provide a theoretical prediction for the spatial variation of the EM spectrum in our Galaxy, which it will be possible to compare to other predictions, based on very different astrophysical models \cite{PorterStrong2005,Maciel2013}. 

\paragraph{Acknowledgements.} I am grateful to Christian Boily and to Garrelt Mellema for useful remarks at the Computational Astrophysics Conference in Saint Petersburg, September 2019.\\




\begin{thebibliography}{9}
\small

\bibitem{BeckWielebinski2013} Beck R, Wielebinski R. Magnetic fields in the Milky Way and in galaxies. In: Planets, Stars and Stellar Systems. Oswalt TD, Gilmore G, editors,  vol. 5, Dordrecht: Springer, 2013, pp. 641--723

\bibitem{Chamandy-et-al2013}
Chamandy L, Subramanian K, Shukurov A. Galactic spiral patterns and dynamo action I: a new twist on magnetic arms. Mon. Not. R. Astron. Soc. 2013;428:3569--3589.

\bibitem{PorterStrong2005} Porter TA, Strong AW. A new estimate of the galactic interstellar radiation field between $ 0.1\mu$m and $1000 \mu$m. In: Proc. 29th International Cosmic Ray Conference, Pune. Mumbai: Tata Institute of Fundamental Research; 2005; Vol. 4, pp. 77-80.

\bibitem{Maciel2013} Maciel WJ.  The interstellar radiation field. In: Astrophysics of the interstellar medium. Maciel WJ, editor. New York: Springer; 2013; Chapter 2, pp. 17-31.

\bibitem{A57}
Arminjon M. On the equations of electrodynamics in a flat or a curved spacetime and a possible interaction energy. Open Physics 2018;16:488--498.

\bibitem{B41}
Arminjon M. Interaction energy of a charged medium and its EM field in a curved spacetime. In: Geometry, integrability and quantization XX; Mladenov IM, Pulov V, Yoshioka A, editors, Sofia: Avangard Prima; 2019; pp. 88-98.

\bibitem{A54} 
Arminjon M. Continuum dynamics and the electromagnetic field in the scalar ether theory of gravitation. Open Physics 2016;14:395--409. 

\bibitem{KentDameFazio1991} Kent SM, Dame TM, Fazio G. Galactic structure from the Spacelab infrared telescope. II. Luminosity models of the Milky Way. Astroph. J. 1991;378:131--138.

\bibitem{Robin-et-al-1992} Robin AC, Cr\'ez\'e M, Mohan V. The radial structure of the Galactic disc. Astron. Astrophys. 1992;265:32--39.

\bibitem{Porcel-et-al-1998} Porcel C, Garz\'on F, Jim\'enez-Vicente J. The radial scale length of the Milky Way. Astron. Astrophys. 1998;330:136--138.

\bibitem{Schneider2006} Schneider P. Extragalactic astronomy and cosmology: an introduction. Berlin: Springer; p. 55.

\bibitem{ZR_et_al2008} Zamboni-Rached M, Recami E, Hern\'andez-Figueroa H E. Structure of nondiffracting waves and some interesting applications. In: Localized waves; Hern\'andez-Figueroa HE, Zamboni-Rached, Recami E, editors. Localized waves. Hoboken: John Wiley \& Sons;  2008; pp. 43--77.

\bibitem{GAZR2014} Garay-Avenda\~no RL, Zamboni-Rached M. Exact analytic solutions of Maxwell's equations describing propagating nonparaxial electromagnetic beams. Appl. Opt. 2014;53:4524--4531.

\bibitem{A60} Arminjon M. An explicit representation for the axisymmetric solutions of the free Maxwell equations. Open Physics 2020;18:255--263.

\bibitem{Jackson1998a}
Jackson JD. Classical electrodynamics. 3rd ed. Hoboken: John Wiley \& Sons; 1998; p. 360.

\bibitem{Durnin1987} Durnin J. Exact solutions for nondiffracting beams. I. The scalar theory. J. Opt. Soc. Am. A 1987;4:651--654.

\bibitem{GradshteynRyzhik2007} 
Gradshteyn IS, Ryzhik IM. Table of Integrals, Series, and Products. 7th English Edition. Burlington (Mass., USA): Academic Press; 2007; \S 6.677, p. 722.


\bibitem{Ala-et-al2011}
Ala G, Francomano E, Viola F. A wavelet operator on the interval in solving Maxwell's equations. Prog.  Electromag. Res. Lett. 2011;27:133--140.


\bibitem{Atkinson1989}
Atkinson KE. An Introduction to Numerical Analysis. 2nd Edition. New York: John Wiley \& Sons; 1989; pp. 257--258.


\end{thebibliography}
\end{document}